\documentclass[iop]{emulateapj}
\usepackage{epsfig}
\usepackage{apjfonts}
\usepackage{aas_macros}

\usepackage{booktabs}
\usepackage{multirow}

\begin{document}

\title{Test the hypothesis of compact-binary-coalescence origin of fast radio bursts through a multi-messenger approach}
\author{Min-Hao Wang$^{1}$, Shun-Ke Ai$^{2}$, Zheng-Xiang Li$^{1}$, Nan Xing$^{1}$, He Gao$^{1,*}$ and Bing Zhang$^{2}$}
\affiliation{
$^1$Department of Astronomy, Beijing Normal University, Beijing 100875, China; gaohe@bnu.edu.cn\\
  $^2$Department of Physics and Astronomy, University of Nevada Las Vegas, NV 89154, USA.}

\begin{abstract}

In the literature, compact binary coalescences (CBCs) have been proposed as one of the main scenarios to explain the origin of some non-repeating fast radio bursts (FRBs). The large discrepancy between the FRB and CBC event rate densities suggest their associations, if any, should only apply at most for a small fraction of FRBs. 
Through a Bayesian estimation method, we show how a statistical analysis of the coincident associations of FRBs with CBC  gravitational wave (GW) events may test the hypothesis of these associations.
We show that during the operation period of advanced LIGO, the detection of $\sim100$ ($\sim1000$) GW-less FRBs with dispersion measure (DM) values smaller than 500 ${\rm pc~cm^{-3}}$ could reach the constraint that less than $10\%$ (or $1\%$) FRBs are related to binary black hole (BBH) mergers. The same number of FRBs with DM values smaller than 100 ${\rm pc~cm^{-3}}$ is required to reach the same constraint for binary neutron star (BNS) mergers. With the upgrade of GW detectors, the same constraints for BBH and BNS mergers can be reached with less FRBs or looser requirements for the DM values. 
It is also possible to pose constraints on the fraction of each type of CBCs that are able to produce observable FRBs based on the event density of FRBs and CBCs. This would further constrain the dimensionless charge of black holes in binary BH systems.

\end{abstract}

	

\section{INTRODUCTION}

Fast radio bursts (FRBs) are bright, milliseconds-duration radio transients with high dispersion measures, typically with an isotropic energy in the radio band as high as $10^{38}$ - $10^{40}$ ergs \citep{Lorimer2007,Thornton2013}. The event rate density of FRBs is about $10^{3}$ to $10^{4} {\rm Gpc^{-3}~yr^{-1}}$ depending on the minimum fluence of the detected FRBs \citep{Petroff2019,CordesandChatterjee2019}.

Even though a growing population of FRBs are found to repeat \citep{Spitler2016,CHIME}, 
the majority of FRBs detected so far are apparently non-repeating. It is possible that a small fraction of FRBs are genuinely non-repeating, which may be associated with catastrophic events.

Many different models have been proposed to explain FRBs, such as binary neutron star mergers \citep{Totani2013,Wang2016,Yamasaki2017,DokuchaevandEroshenko2017}, binary white-dwarf mergers \citep{Kashiyama2013}, mergers of charged black holes \citep{Zhang2016b,Liu2016}, collapses of supramassive rotating neutron stars \citep{FalckeandRezzolla2013,Zhang2014,Ravi2014,PunslyandBini2016}, magnetar flares \citep{PopovandPostnov2010,Kulkarni2014,Lyubarsky2014}, black hole batteries \citep{Mingarelli2015}, collisions and interactions between neutron stars and small objects \citep{GengandHuang2015,HuangandGeng2016,Dai2016,MottezandZarka2014,Smallwood2019}, quark novae \citep{Shand2016}, giant pulses of pulsars \citep{Connor2016,CordesandWasserman2016}, cosmic combs \citep{Zhang2017,Zhang2018}, superconducting cosmic strings \citep{Yu2014}. See \cite{platts18} for a review on the available theoretical models. 

A good fraction of these models are related to compact binary coalescences (CBCs), including binary neutron star (BNS) mergers, binary black hole (BBH) mergers and black hole-neutron star (BH-NS) mergers. For BNS mergers, there have been several proposals.
\cite{Totani2013} suggested that synchronization of the magnetosphere of the two NSs shortly after the merger can power bright coherent radio emission in a manner similar to radio pulsars. \cite{Zhang2014} suggested that if the BNS merger product is a supramassive NS \citep{Dai2006,zhang13,gao16}, an FRB can be produced as the supramassive NS collapses into a black hole as the magnetic ``hair'' of the black hole is ejected \citep{FalckeandRezzolla2013}. \cite{Wang2016} proposed that during the final inspiral phase, an electromotive force would be induced on one NS to accelerate electrons to an ultra-relativistic speed instantaneously, thus generate FRB signals via coherent curvature radiation from these electrons moving along magnetic field lines in the magnetosphere of the other NS. So, theoretically, an FRB can accompany a BNS merger event right before \citep{Wang2016}, during \citep{Totani2013} or 100s of seconds after \citep{Zhang2014,Ravi2014} the merger. 
For BBH and plunging BH-NS (mass ratio less than 0.2 \citep{shibata2009}) mergers, one would not expect bright electromagnetic counterparts for CBCs. However, if at least one of the members is charged, both dipole electric radiation and dipole magnetic radiation would be emitted from the system during the inspiral phase. The emission powers increase sharply at the final phase of the coalescence \citep{Zhang2016b,Zhang2019,Deng2018}. This would produce a brief electromagnetic signal, which may manifest itself as an FRB if coherent radio emission can be produced from the global magnetosphere of the system \citep{Zhang2016b,Zhang2019}.

The host galaxy information is helpful to constrain the origin of FRBs. The first repeating FRB 121102 was localized in a dwarf galaxy with a redshift of 0.19273 \citep{Spitler2016,Scholz2016,Chatterjeeet2017,Marcote2017,Tendulkar2017}. Most recently, two non-repeating FRBs were  precisely localized (FRB 180924 \citep{Bannister2019}, FRB 190523 \citep{Ravi2019}). Interestingly, unlike FRB 121102, the host galaxies of the latter two apparently non-repeating FRBs have a relatively low star-formation rate. The locations of the FRBs have a relatively large spatial offset with respect to the host galaxy \citep{Bannister2019,Ravi2019}. These properties are similar to those of short GRBs believed to be produced by neutron star mergers \citep{Berger2013}. These discoveries therefore revive the possibility that a fraction of FRBs might be related to binary neutron star (BNS) or neutron star-black hole (NS-BH) mergers. Since the FRB event rate density is much higher than those of CBCs and since a good fraction of FRBs repeat, the CBC-associated FRBs, if exist, should only comprise of a small fraction of the full FRB population. 

CBCs are the sources of gravitational waves (GWs). A direct proof of the CBC-related FRBs would be the direct observation of FRB - CBC associations. So far, no such associations have been found. The non-detection could be discussed in two different contents. If a CBC is detected without an associated FRB counterpart, one may not draw firm conclusions regarding the non-associations. This is because current radio telescopes to detect FRBs do not cover the all sky, so that one cannot rule out the existence of an associated FRB with the CBC. Even if the entire CBC error box was by chance covered by radio telescopes, one cannot rule out the association since a putative FRB might be beamed away from Earth. On the other hand, if a FRB is detected without an associated GW signal, the constraints on the association would be much more straightforward. First, the FRB source may be outside the GW detection horizons. If one only focuses on those FRBs that are within the horizons of GW detectors, the non-detection of an association only has one possible reason: the FRB is not from a CBC. By observing many of such FRBs, one would be able to constrain the fraction $f$ of CBC-origin FRBs.

In this paper we develop a Bayesian model to estimate the fraction $f$ based on the joint (non)-detection of FRBs and GWs. We claim that even for GW-less FRBs (FRBs  without detected GW counterparts), a accumulation of the sample can place a constraint on $f$. Furthermore, based on the event rates of FRBs and BBH mergers, one may also constrain the charge of the black holes in the BBH and/or NS-BH systems.

\section{Methods}
\subsection{Bayesian estimation Model}
Suppose that during the all-sky monitoring of CBC events by GW detectors a sample of FRBs are detected, which could be denoted as $D=(D_1,D_2,...,D_N)$, where $N$ is the total number of the FRBs in the sample. One can define $D_i=(d_i,{\rm DM}_i)$, where ${\rm DM}_i$ is the DM value for the \textit{i}th FRB, and $d_i$ represents whether the \textit{i}th FRB is detected ($d_i=1$) by the GW detectors or not ($d_i=0$). For DM estimation, basically three components should be considered, only which from the intergalactic medium (IGM) is supposed to depend on the cosmological distance. Besides the IGM component, contributions from the Milky Way (MW), and the FRB host galaxy (host) are also needed to be considered. 

Since ${\rm DM_{MW}}$ and ${\rm DM_{host}}$ can be only roughly modeled by simple distributions, one particular $z$ may correspond to a wide distribution of possible $DM$ values. In other words, a particular DM value may correspond to a wide distribution of $z$. We use $P_i$ to represent the probability of the \textit{i}th FRB being within the detection horizon of the GW detectors ($z_h$, in terms of redshift). If the redshift of the \textit{i}th FRB ($z_i$) could be determined, it is relatively easy to get $P_i=1$ (when $z_i<z_h$) or $P_i=0$ (when $z_i>z_h$).

A Bayesian formula can be used to estimate the probability distribution of $f$ as
\begin{eqnarray}
\pi(f|D)= \frac{L(D;f)\pi({f})}{\int{L(D;f)}\pi({f})df},
\end{eqnarray}
where $\pi({f})$ is the prior distribution for $f$ and $L(D;f)$ represents the likelihood function for observing $D=(D_1,D_2,...,D_N)$ sample under the hypothesis that a fraction $f$ of the FRBs come from a specific kind of CBC events. Here we have 
\begin{eqnarray}
L(D;f)=C_N^m \prod L(D_i;f)=C_N^m \prod [d_i f P_i+(1-d_i)(1-f P_i)], 
\end{eqnarray}
where $m$ is the number of FRBs with GW detections for CBCs and $N$ is the total number of FRBs.

One can apply this model to constrain FRBs from any kind of CBC event. Ignoring the uncertainty of DM models, only the horizon $z_h$ influence the final results, which is determined by both the CBC types and GW detectors.

\subsection{DM models and samples}
To be specific, the observed DM value could be expressed as 
\begin{equation}\label{eqdmobs}
\rm{DM}_{\rm{obs}}=\rm{DM}_{\rm{MW}}+\rm{DM}_{\rm{IGM}}+\rm{DM}_{\rm{host}}.
\end{equation}
${\rm DM}_{\rm IGM}$ depends on the cosmological distance scale and the fraction of ionized electrons in hydrogen (H, $\chi_{\rm{e,H}}(z)$) and helium (He, $\chi_{\rm{e,He}}(z)$) along the path. The latter two elements are closely related to the present-day baryon density parameter $\Omega_\mathrm{b}$ and the fraction of baryons in the IGM, $f_{\rm{IGM}}$. If both hydrogen and helium are fully ionized (valid below $z\sim3$), the average value (for individual line of sight, the value may deviate from this due to the large scale density fluctuations, \citealt{mcquinn14}) can be written as \citep{Gao2014}
\begin{equation}\label{eqdmigm}
\mathrm{DM}_{\mathrm{IGM}}(z)=\frac{21cH_0\Omega_bf_{\mathrm{IGM}}}{64\pi G m_{\mathrm{p}}}\int_0^z\frac{(1+z')dz'}{E(z')}.
\end{equation} 
The uncertainty of $\rm DM_{IGM}$ is important but complicated because of the density fluctuation from the large scale structure. According to \citet{mcquinn14}, the standard deviation from the mean DM is dependent on the profile models characterizing the inhomogeneity of the baryon matter in the IGM. Here, we use numerical simulation results of  \citet{mcquinn14} and \citet{faucher-giguere11} (purple dotted line in the bottom panel of Fig. 1 in  \citet{mcquinn14}) to account for the standard deviation. 

Here, DM contribution from the Milky Way is derived by modeling the electron density distribution in a spiral galaxy with the NE2001 model and considering a uniform spatial distribution of FRBs~\citep{CordesandLazio2002,XuHan2015}.  
The value of $\rm {DM_{host}}$ and its uncertainty $\sigma_{\rm{host}}$ are intractable parameters since they are poorly known and related to many factors, such as the local near-source plasma environment, the site of FRB in the host, the inclination angle of the galaxy disk, and the type of the host galaxy \citep[e.g.][]{XuHan2015,luo18}. In our analysis, we assume that the type of the host galaxy is similar to the one of the Milky Way. Moreover, an additional contribution from the local nearby plasma also should be taken into account. Here, we use $\rm {DM_{host}}$ to denote the total contribution from both the host galaxy and the local nearby environment. For an FRB at redshift $z$, the rest-frame $\rm {DM_{host}}$ relates to the contribution to the observed DM via a factor $1+z$, i.e. $\rm {DM_{host}}=\rm {DM_{host,loc}}/(1+z)$.

With all three budgets in Eq. \ref{eqdmobs} addressed, we mock a sample containing $\sim 10^6$ ($10^7$) FRBs with the redshift uniformly distributed in $z = 0 - 1$ ($0 - 9$). In our simulation, we assume $f_{\rm  IGM}=0.83$ and a \citet{planck18} cosmology with $\Omega_m=0.3153,~\Omega_bh^2=0.0224$, and $h=H_0/100~{\rm km~s^{-1}~Mpc^{-1}}=0.6736$. Based on our simulated sample, $P_i$ could be estimated for any given ${\rm DM}_i$ and $z_h$. 

\section{constrain the fraction of FRB\lowercase{s} from CBC\lowercase{s}}
To constrain the fraction of FRBs from different kinds of CBC events, the horizon of the GW detector is a key parameter. In principle, the GW horizon of each kind of CBC event is a function of mass of the system. Here we choose some characteristic masses for different types of CBCs as an example.
 
For NS-NS mergers, the horizon is $\sim 220{\rm Mpc}$ ($z_h \approx 0.05$) for aLIGO \citep{abramovici92}, $480{\rm Mpc}$ ($z_h\approx 0.1$) for aLIGO A+ \citep{ligo16} and $\sim 2.3{\rm Gpc}$ ($z_h\approx 0.5$) for the proposed third generation GW detector Einstein Telescope (ET) \citep{et17}; For BH-BH mergers with a total mass of $\sim 60M_{\odot}$ ($30M_{\odot}+30M_{\odot}$), the horizon  is $\sim 1.6{\rm Gpc}$ ($z_h \approx 0.3$) for aLIGO, $2.5{\rm Gpc}$ ($z_h\approx 0.45$) for aLIGO A+ and $\sim 354{\rm Gpc}$ ($z_h\approx 40$) for ET \citep{ligo19}.

\begin{figure}[ht!]
\resizebox{90mm}{!}{\includegraphics[]{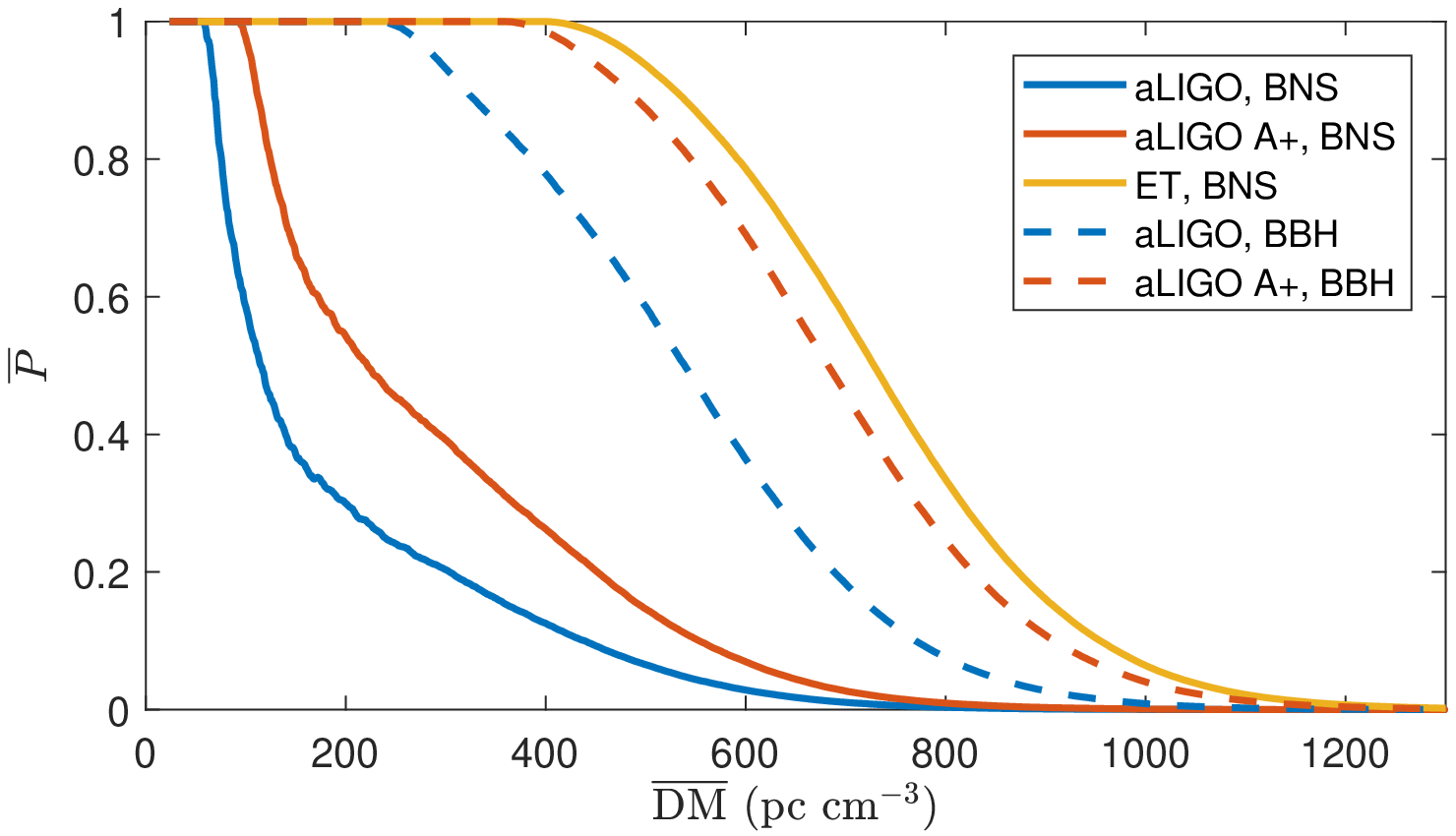}} \\
\resizebox{90mm}{!}{\includegraphics[]{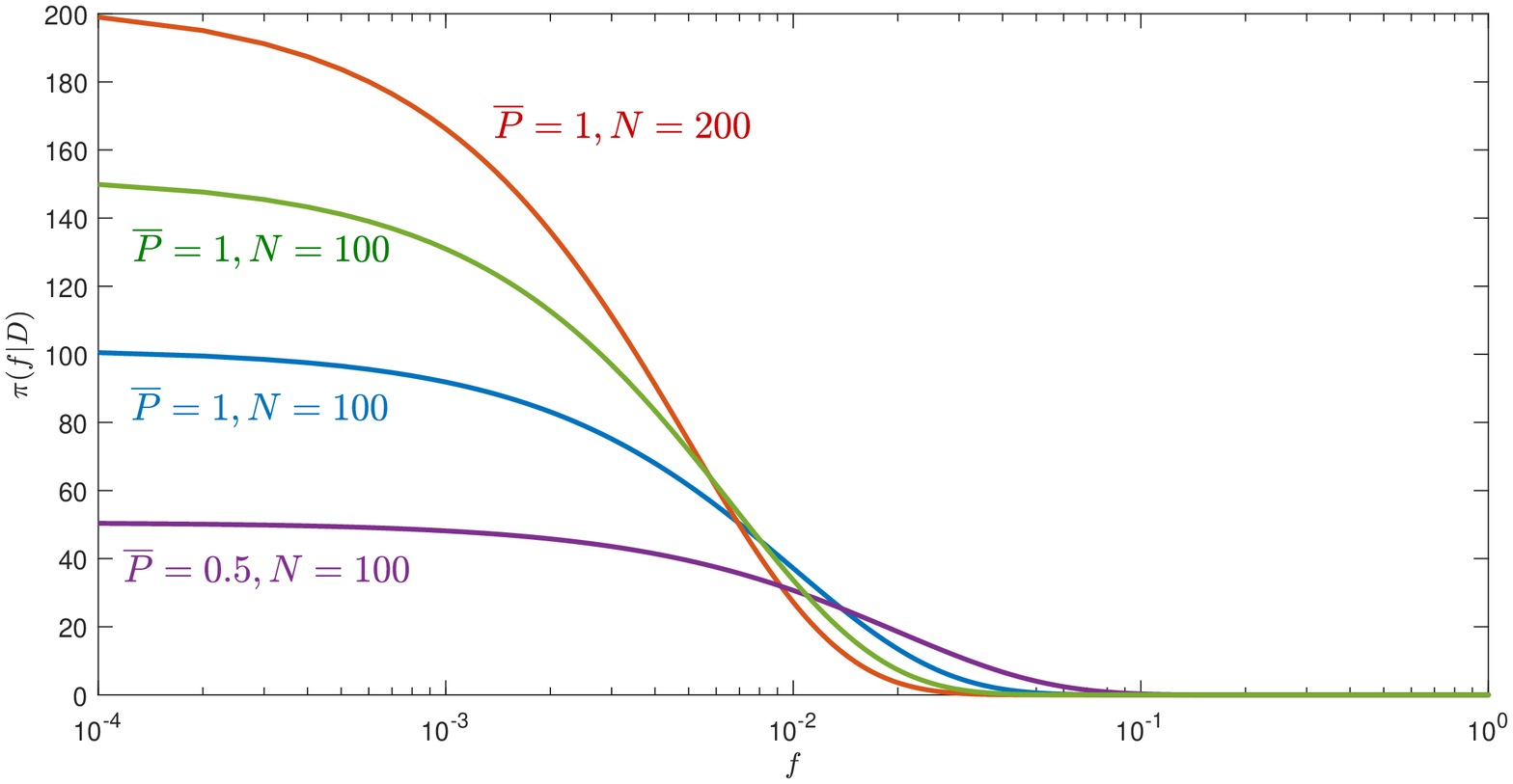}} 
\caption{The upper panel shows the possibility $\bar P$ of FRBs with given DM values locating within the horizon of different GW detectors. Solid lines are for BNS cases and dashed lines are for BBH cases. Lines with different colors refer to different GW detectors. The lower panel shows the posterior distribution of the fraction $f$ after $N$ FRBs with the same DM value (so is the $\bar P$ value) being detected. Lines with different colors corresponds to different values of $N$ and $\bar P$.}
\label{distribution}
\end{figure}

\begin{figure*}[ht!]
\begin{center}
\begin{tabular}{lll}
\resizebox{90mm}{!}{\includegraphics[]{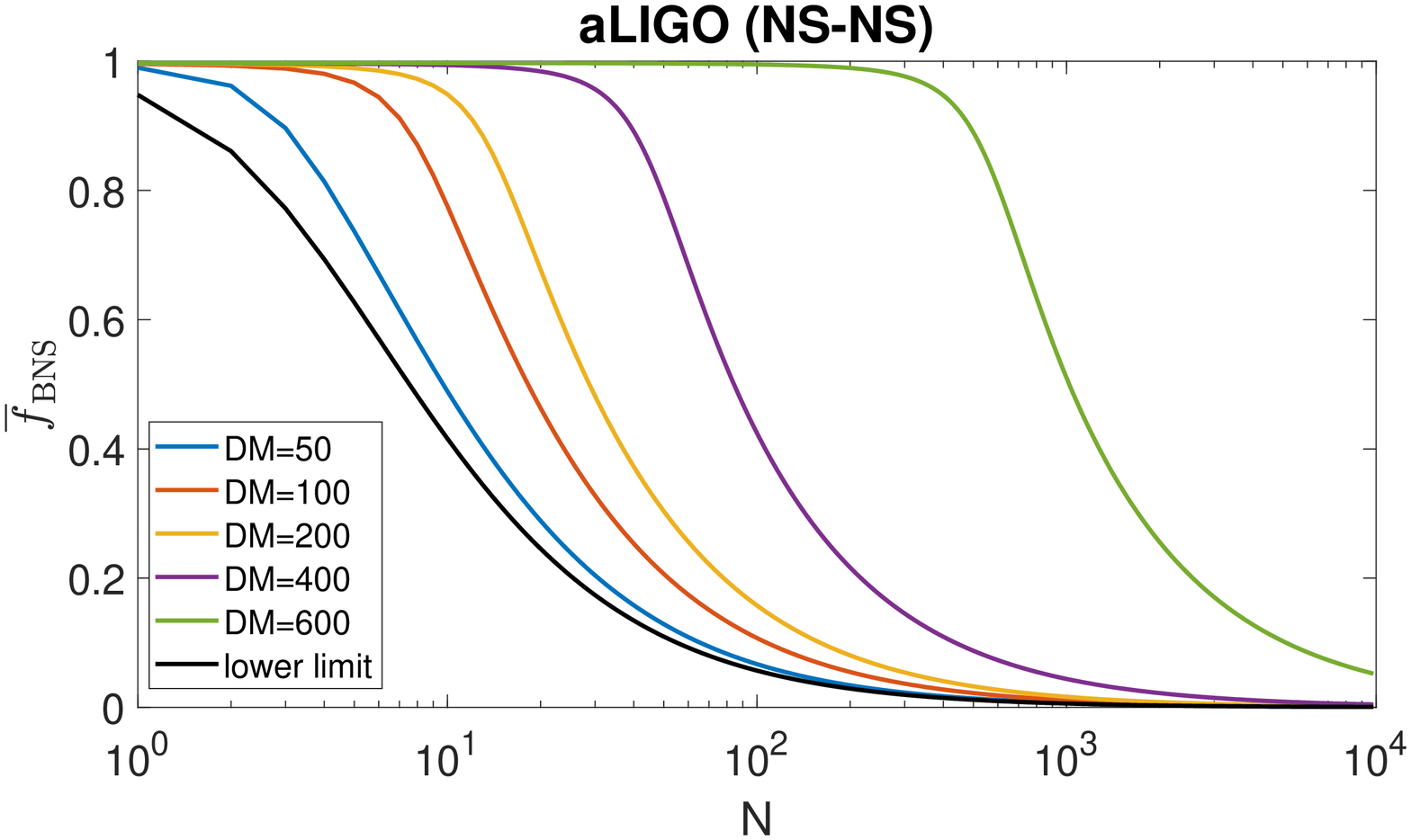}} &
\resizebox{90mm}{!}{\includegraphics[]{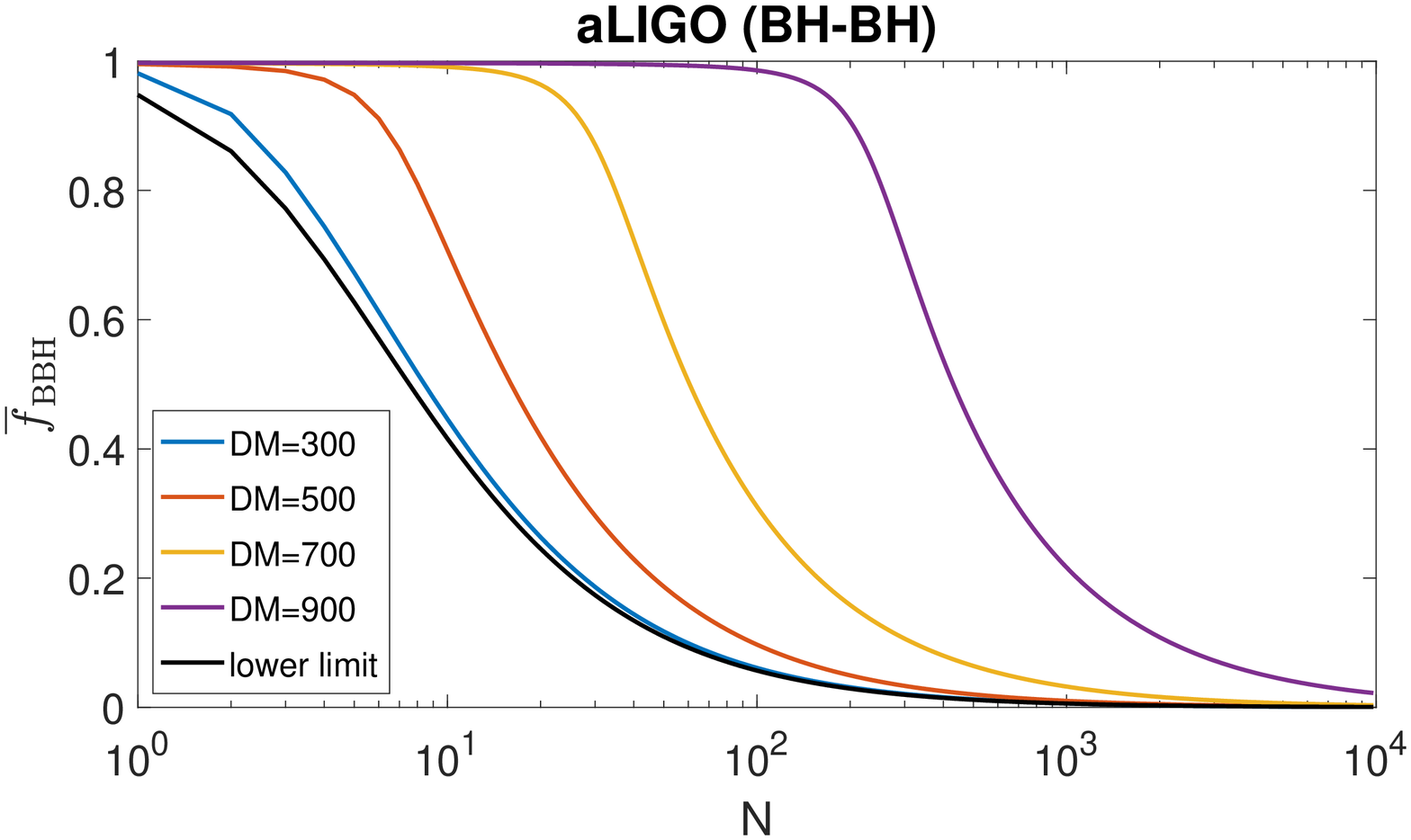}} \\
\resizebox{90mm}{!}{\includegraphics[]{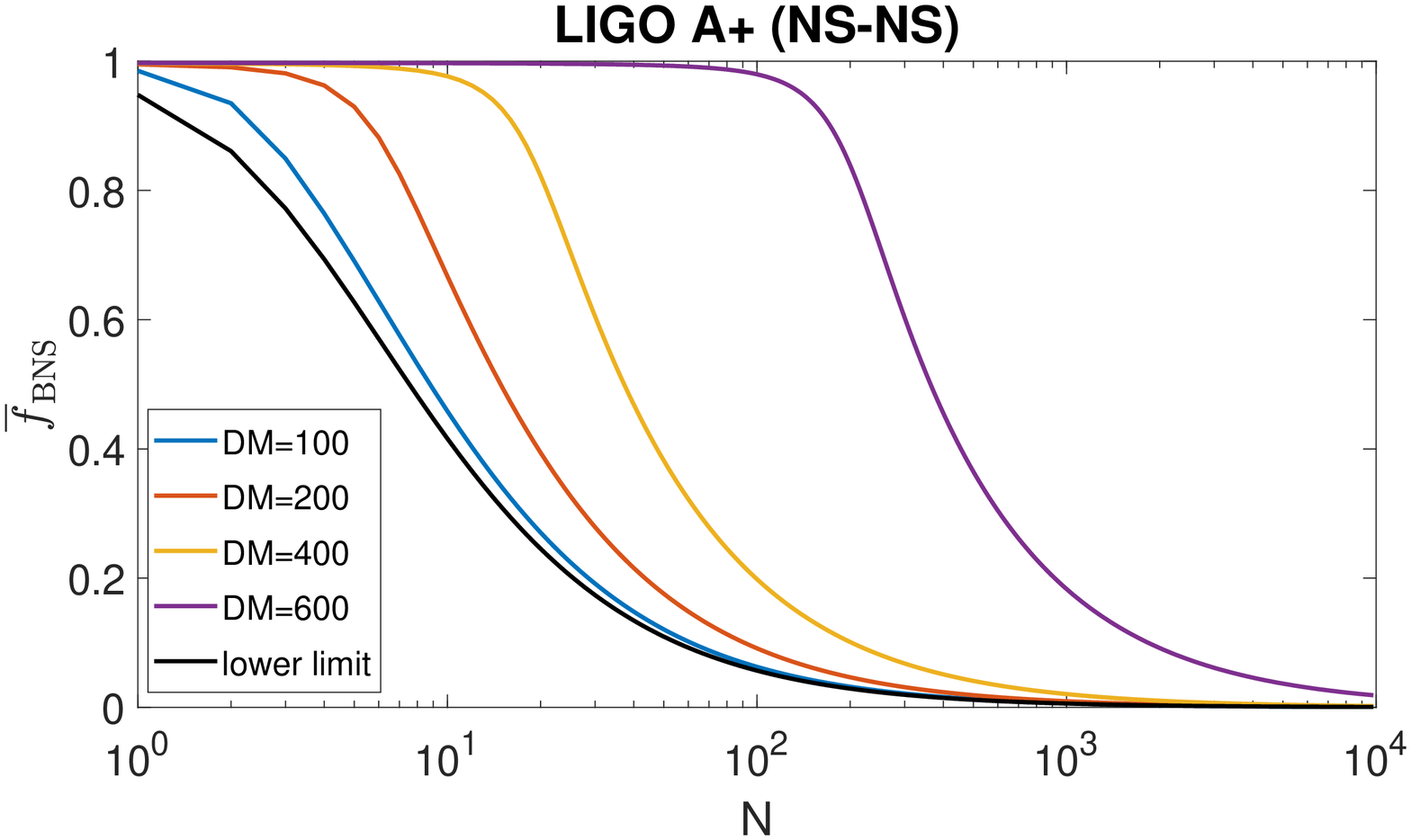}} & 
\resizebox{90mm}{!}{\includegraphics[]{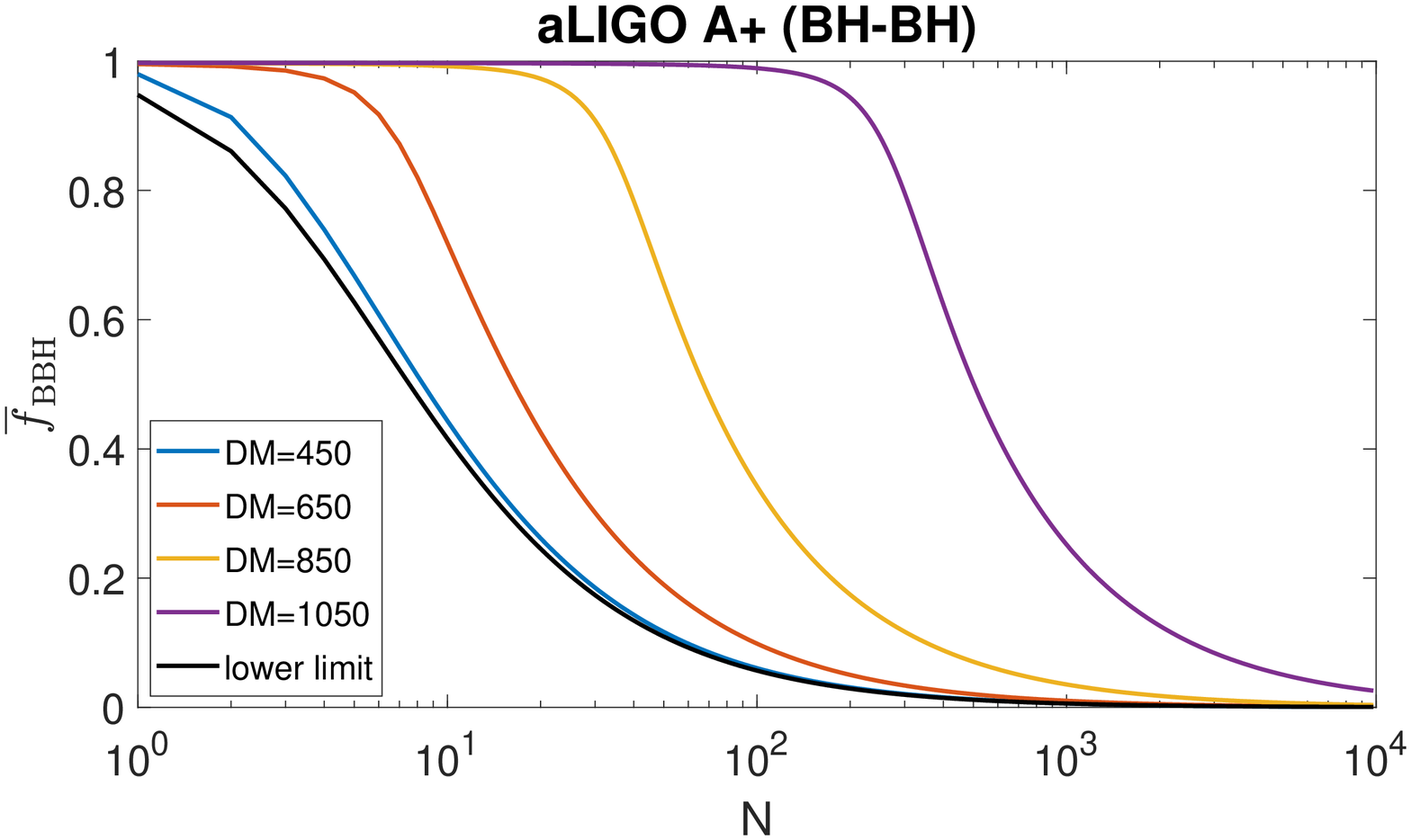}}\\
\resizebox{90mm}{!}{\includegraphics[]{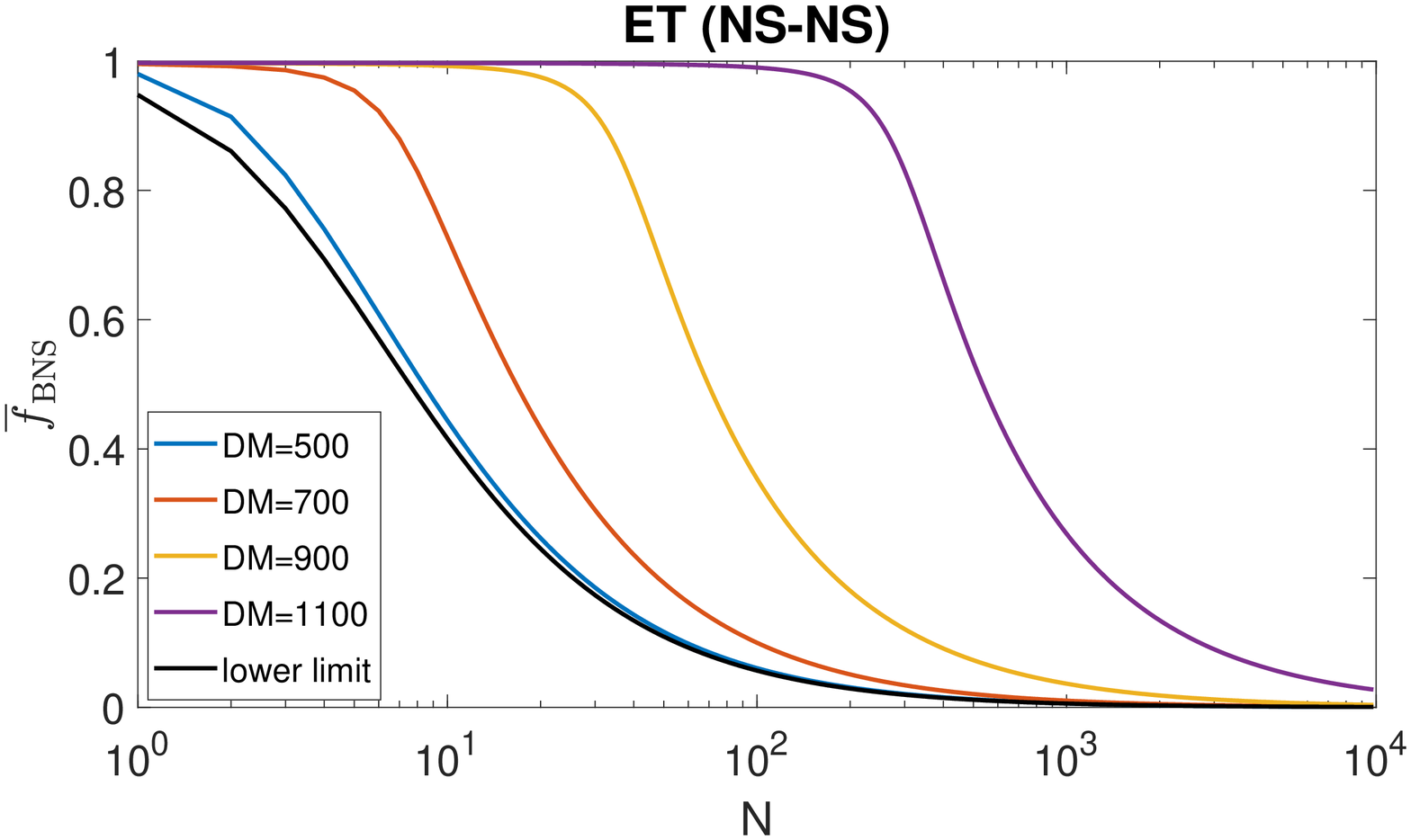}}  \\
\end{tabular}
\caption{Constraints on $\overline{f}$ as the FRB sample with different $\overline{DM}$ values accumulating for various GW detectors operation. The left panel represents the constraints from FRBs and GW observations of BNS merger. The right panel represents the constraints from FRBs and GW observations of BBH merger. The black line in each figure stands for the case that FRB sources are always supposed to within the horizon of GW detector.} 
\label{constraint}
\end{center}
\end{figure*}

\begin{table}[!hbp]
\centering
\caption{Constraints on $\overline{f}$ for different size of the FRB samples with different $\overline{DM}$ values under various GW detectors operation.}
\begin{tabular}{c|ccc|ccc|ccc}
\toprule  
&\multicolumn{3}{ c }{aLIGO}& \multicolumn{3}{ c }{LIGO $A+$}& \multicolumn{3}{ c }{ET}\\
\midrule 
&$\overline{DM}$& $\overline{f}_{\rm BNS}$& N& $\overline{DM}$& $\overline{f}_{\rm BNS}$& N& $\overline{DM}$& $\overline{f}_{\rm BNS}$& N\\
\midrule  
\multirow{15}{*}{NS-NS}&50& 50\%& 10& 100& 50\%& 9& 500& 50\%& 9\\ 
&50& 10\%& 65& 100& 10\%& 62& 500& 10\%& 59\\ 
&50& 5\%&135&100& 5\%&127& 500& 5\%& 122\\ 
&50& 1\%& 692& 100& 1\%& 648& 500& 1\%& 628\\
&100& 50\%& 18& 200& 50\%& 15& 700& 50\%& 17\\ 
&100& 10\%& 108& 200& 10\%& 91& 700& 10\%& 100\\ 
&100& 5\%&220& 200& 5\%&186& 700& 5\%& 204\\
&100& 1.1\%& 1000& 200& 1\%& 946& 700& 1\%& 1000\\ 
&200& 50\%& 29& 400& 50\%& 37& 900& 50\%& 70\\
&200& 10\%& 160& 400& 10\%& 202& 900& 10\%& 364\\
&200& 5\%& 325& 400& 5\%&409& 900& 5\%& 733\\
&200& 1.6\%& 1000& 400& 2.1\%& 1000& 900& 3.7\%& 1000\\
&400& 50\%& 84   & 600& 50\%& 364& 1100& 50\%& 535\\
&400& 10\%& 438&   600& 18\%&1000& 1100 &27\%& 1000\\
&400& 5\%& 880&   &    &    &    &    & \\
&600& 51\%& 1000&  &    &    &    &    & \\
\midrule 
&$\overline{DM}$& $\overline{f}_{\rm BBH}$& N& $\overline{DM}$& $\overline{f}_{\rm BBH}$& N& $\overline{DM}$& $\overline{f}_{\rm BBH}$& N\\
\midrule  
\multirow{14}{*}{BH-BH}&300& 50\%& 9& 450& 50\%& 8&-- &50\% & 8\\ 
&300& 10\%& 60& 450& 10\%& 59&-- &10\% & 55 \\ 
&300& 1\%& 631& 450& 1\%& 627&-- &1\% & 590\\ 
&500& 50\%& 16& 650& 50\%& 16& & & \\ 
&500& 10\%& 97& 650& 10\%& 99& & & \\ 
&500& 1\%& 1000& 650& 1\%& 1000& & & \\  
&700& 50\%& 61&  850& 50\%& 67& & & \\
&700& 10\%& 319& 850& 10\%& 352& & & \\
&700& 3.2\%& 1000& 850& 1\%& 3556& & & \\
&900& 50\%& 431& 1050& 50\%& 502& & & \\
&900& 10\%& 2193& 1050& 10\%&2518& & & \\
&900& 2.2\%& 10000& 1050& 2.5\%&10000& & & \\

\bottomrule 
\end{tabular}
\label{results}
\end{table}

For a specific GW detector, one can use our proposed Bayesian estimation model to calculate the posterior probability density distribution of $f$ for a given FRB sample $D=(D_1,D_2,...,D_N)$, that may be detected in the future. 
As an example, here we focus on the accumulation of negative joint detection case, which means a large sample of FRBs are detected during the GW detector operation but have no joint GW signals detected, so that $m=0$ and $d_{i=1...N}=0$. For simplicity, we assign a characteristic DM value for the whole sample, namely $DM_{i=1...N}=\overline{DM}$. Since only a small fraction of FRBs are expected to be well localized, which is at least true in the near future, here we assume that not all $z_i$ could be well determined and  all the $P_i$'s are estimated with the Monte Carlo simulation method. Similarly, $P_{i=1...N}=\overline{P}$ is assumed. As shown in figure \ref{distribution}, $\overline{P}$ decreases from 1 to 0 with the increase of $\overline{DM}$, because FRBs with smaller DM values are more likely to be within the horizon of GW detectors. Based on such a mock observational FRB sample, $\overline{P}$ can be calculated, so is the posterior probability density distribution of $f$.
The results are shown in Figure \ref{distribution}. Note that we have taken the prior distribution of $f$ as a uniform distribution. 

Since so far no detected FRBs are accompanied with GW triggers, the posterior probability density distribution of $f$ peaks at $f=0$. Given the value of $\overline{DM}$ and $z_h$, the posterior probability density distribution of $f$ would become narrower as the sample accumulates, whereas given the sample size $N$, the distribution would become narrower as $\overline{DM}$ decreases or $z_h$ increases. Here we define $\overline{f}$ as the upper limit of the fraction of FRBs being associated with a specific type of CBCs, where the probability of  $f<\overline{f}$ is larger than $99.7\%$ (equivalent 3$\sigma$ confidence level). In Figure \ref{constraint} and Table 1, we show how $\overline{f}$ evolves as the sample accumulates for different $\overline{DM}$ values and different GW detectors. 

It is obvious that when FRBs with small $\overline{DM}$ values are considered, which means all of the FRB sources are supposed to be within the horizon of GW detectors, only a small number of FRB detections without GW counterparts can lead to a low level of $\overline{f}$. This case is shown with black lines in Figure \ref{constraint}. To be specific, $\sim 10$ FRBs without GWs can constrain $\overline{f}$ below $50\%$; $\sim 55$ FRBs without GWs can constrain $\overline{f}$ below $10\%$; $\sim 590$ FRBs can constrain $\overline{f}$ below $1\%$. 

As shown in Table \ref{results}, for a certain GW detector toward a specific type of CBCs, the increase of $\overline{DM}$ lead to looser constraints. In other words, more detections are required to obtain the same constraint on $\overline{f}$. However, for different GW detectors, to reach the same constraint level with the same number of detections, the required $\overline{DM}$ is totally dependent on the horizon of the GW detectors.

From GW observations, the event rate density of BBH mergers and BNS mergers are estimated as \citep{Abbott2019,LIGO2020}
\begin{eqnarray}
\dot{\rho}_{\rm BBH}\sim 53.2^{+58.5}_{-28.8}{\rm Gpc^{-3} yr^{-1}},
\end{eqnarray}
with $90\%$ confidence level and 
\begin{eqnarray}
\dot{\rho}_{\rm BNS}\sim 250-2810{\rm Gpc^{-3} yr^{-1}},
\end{eqnarray}
which are obviously lower than that of FRBs, which could be estimated as\footnote{The estimation is good for FRBs with luminosity larger than $10^{43}~\rm ergs^{-1}$. If a significant fraction of FRBs have lower luminosity, the FRB event rate could be even larger. In this case, the maximum possible value of $\overline{f}_{\rm BBH}$ and $\overline{f}_{\rm BNS}$ would be even smaller, so that more GW-less FRBs are needed to achieve meaningful constraints with our proposed method.} \citep{Zhang2016b}
\begin{eqnarray}
\dot{\rho}_{\rm FRB}\sim (5.7 \times 10^{3} {\rm Gpc^{-3} yr^{-1}}) \times \left({D_z \over 3.4{\rm Gpc}}\right)^{-3}\left({\dot{N}_{\rm FRB} \over 2500}\right).
\end{eqnarray}
Here the all-sky FRB rate $\dot{N}_{\rm FRB}$ is normalized to $2500 \ {\rm d}^{-1}$ \citep{Keane2015}, and the comoving distance $D_z$ is normalized to $z=1$. The ratio between the rates of different kinds of CBCs and the event rate of FRBs provides the maximum possible value of $\overline{f}$. Based on current results, we have $\overline{f}_{\rm BBH}<0.93^{+1.03}_{-0.50}\%$ (with $90\%$ confidence level) and $\overline{f}_{\rm BNS}<4.39\%-49.3\%$. According to Table 1, we find that for aLIGO (LIGO A+), $\sim 1000$ GW-less FRBs with $\overline{DM}< 500 ~(600) {\rm pc~cm^{-3}}$ could achieve a meaningful constraint, where $\overline{f_{\rm BBH}}<1\%$, while for the third generation of GW detector ET, almost all the sources of FRBs are within its horizon for BBH mergers, so the constraints come to the limiting case shown with the black lines in Figure \ref{constraint}: $\sim 600$ FRBs with arbitrary $\overline{DM}$ value can reach the constraint that less than $1\%$ FRBs are related to BBH mergers. There is a big uncertainty for BNS merger rates, so is the maximum possible value of $\overline{f}_{\rm BNS}$. In an optimistic situation ($\overline{f}_{\rm BNS}<49.3\%$), we find that for aLIGO (LIGO A+), $\sim 30 (15)$ GW-less FRBs with $\overline{DM}< 200 {\rm pc~cm^{-3}}$ could achieve a meaningful constraint, where $\overline{f_{\rm BNS}}<50\%$, and $\sim 1000 (400)$ GW-less FRBs with $\overline{DM}< 600 {\rm pc~cm^{-3}}$ could reach the same constraint. For ET,  $\sim 10$ FRBs with $\overline{DM}< 500 {\rm pc~cm^{-3}}$ can reach the constraint that less than $50\%$ FRBs are related to BNS mergers. On the other hand, in a pessimistic situation ($\overline{f}_{\rm BNS}<4.39\%$), we find that for aLIGO (LIGO A+), $\sim 400 (200)$ GW-less FRBs with $\overline{DM}< 200 {\rm pc~cm^{-3}}$ could achieve a meaningful constraint, where $\overline{f_{\rm BNS}}<5\%$, and $\sim 1000 (500)$ GW-less FRBs with $\overline{DM}< 400 {\rm pc~cm^{-3}}$ could reach the same constraint. For ET,  $\sim 140$ FRBs with $\overline{DM}< 500 {\rm pc~cm^{-3}}$ can reach the constraint that less than $5\%$ FRBs are related to BNS mergers. It is interesting to note that, in this case, for a similar $\overline{DM}$ value and a same GW detector, the required sample size of FRBs is comparable between BNSs and BBHs, in order to achieve meaningful constraints. 

Note that here we only show results for BNS and BBH, since the constraints for NS-BH merger model should be similar with the BNS merger case, except that the horizon of GW detectors for NS-BH mergers is slightly larger than that for BNS mergers, which leads to a more stringent constraint on $\overline{f}_{\rm NS-BH}$ with the same DM values and number of detections. The example we show here is based on a simplified situation that a characteristic DM value is assigned for the entire FRB sample, and that all FRBs in the sample are neither well localized nor associated with a GW detection. The results could be used as a reference for more realistic cases. For instance, if we have an FRB sample with a characteristic DM value as the maximum of the whole sample, namely $DM_{i=1...N}\leq\overline{DM}$, in order to achieve a similar constraint on $\overline{f}$, much less FRBs are required, i.e. $N$ value in Table 1 would become much smaller. On the other hand, if some precise positioning is achieved for some FRBs in the sample, and if their distances are determined within the detection horizon of the monitoring GW detectors but there is no GW detection, these sources will increase their weight so that fewer samples are needed to obtain the same constraint on $\overline{f}$. Finally, if some FRBs in the sample are associated with GW signals and the signals are from one kind of CBC events, then the distribution center value of $f$ for this CBC-origin FRB model is no longer 0, but the upper limit of the proportion could still be limited with the accumulation of FRBs in the sample.

\section{Constraints on BH charge}

A number of FRB models based on BNS mergers have been proposed. These models invoke different BNS merger physics, so it is not easy to constrain NS properties through negative joint detection between FRBs and BNS merger GW events. On the other hand, the FRB model based on BBH mergers directly depends on the amount of dimensionless charge carried by the BHs with essentially no dependence on other parameters \citep{Zhang2016b,Zhang2019}. Accumulation of FRBs without BBH merger associations can hence place interesting constraints on the amount of charge carried by BHs.

According to \cite{Zhang2016b}, an FRB may be made from BBH mergers when at least one of the BH carries a dimensionless charge $\hat q \equiv Q/Q_c > 10^{-9}-10^{-8}$, where $Q_c=2\sqrt{G}M=(1.0\times 10^{31}{\rm e.s.u.})({M / 10M_{\odot}})$. Assuming that the radio efficiency of a charged CBC luminosity is $\eta_r$ and equal mass in the BBH system, the FRB luminosity can be estimated as  \citep{Zhang2019}
\begin{eqnarray}
    L_{\rm FRB} & = & \frac{1}{6} \frac{c^5}{G} \hat q^2 \eta_r \left(\frac{r_s}{a}\right)^{-4} = \frac{1}{96} \frac{c^5}{G} \hat q^2 \eta_r \nonumber \\
    & = & (3.8\times 10^{57} \ {\rm erg \ s^{-1}}) \ \xi^2,
\end{eqnarray}
where 
\begin{eqnarray}
   \xi\equiv\hat q \sqrt{\eta_r}\,,
\end{eqnarray}
$r_s=2GM/c^2$ is the Schwarzschild radius of each BH and  $r_s/a = 1/2$ at the merger. 

For a sample of GW-less FRB detection, where the minimum FRB luminosity within the sample is $L_{\rm min}$, one can define a critical value for the combination of BH charge and radio efficiency, $\xi_c$, where 
\begin{eqnarray}
(3.8\times 10^{57} \ {\rm erg \ s^{-1}}) \ \xi_c^2 = L_{\rm min},
\end{eqnarray}
namely
\begin{equation}
    \xi_c^2= 2.6\times 10^{-17} L_{\rm min,41},
\end{equation}
or
\begin{equation}
    \xi_c = 5.1\times 10^{-9} L_{\rm min,41}^{1/2}.
    \label{eq:q-constraints}
\end{equation}
Notice that the FRBs produced by charged BBH mergers are essentially isotropic. If all BBH systems are charged, and a good fraction of BBH systems satisfy $\xi > \xi_c$, with sufficient FRB sample size, there should be some FRBs together with GW counterparts detected. Otherwise, we can put an upper limit to the fraction of BBH systems with $\xi > \xi_c$, which could be estimated as
\begin{eqnarray}
F_{\xi > \xi_c}&=&{\dot{\rho}_{\rm FRB} \over \dot{\rho}_{\rm BBH}}\times \overline{f}_{\rm BBH}\nonumber \\
&\sim& 1.0^{+1.1}_{-0.5}\times \left({D_z \over 3.4{\rm Gpc}}\right)^{-3}\left({\dot{N}_{\rm FRB} \over 2500}\right)\times \left({f_{\rm BBH} \over 0.93\%}\right).
\label{eq:F}
\end{eqnarray}
Here, we normalize $\overline{f}_{\rm BBH}$ to $0.93\%$, which is the maximum possible value of $\overline{f}_{\rm BBH}$ according to current observations. Obviously, a more stringent constraint on $\overline{f}_{\rm BBH}$ leads to a more meaningful constraint on $F_{\xi > \xi_c}$.

\section{CONCLUSION AND DISCUSSION}
Many models have been proposed to explain the origin of FRBs. Among them several CBC-origin models have been discussed to interpret non-repeating FRBs. Since CBCs are main targets for GW detectors, it is possible to combine the joint FRB and GW data to test these hypotheses. 
Since the event rate density of FRBs is much greater than the event rate density of CBCs, it is believed that at most only a small portion of FRBs could originate from CBCs. The continuous observational campaigns in both the GW field and the FRB field makes it possible to achieve FRB-GW joint detections if such associations are indeed realized in Nature. A sufficient number of the non-detections of GW sources from FRBs can also place interesting constraints on these scenarios. We developed a Bayesian estimation method to constrain the fraction $f$ of CBC-origin FRBs using the future joint GW and FRB observational data. 

The size of the FRB sample needed to make a sufficient constraint depends on the GW detection horizon for the type of CBCs in discussion and the DM values of the observed FRBs. 
According to the published FRB sample, the mean value of DM distribution is approximately 668.3${\rm pc~cm^{-3}}$, with the range of 203.1${\rm pc~cm^{-3}}$ to 1111${\rm pc~cm^{-3}}$ for the 1 $\sigma$ confidence interval and 103.5${\rm pc~cm^{-3}}$ to 1982.8${\rm pc~cm^{-3}}$ for the 3 $\sigma$ confidence interval\footnote{Here we use the data presented in the FRB catalogue \cite{Petroff2016} from the url $(http://www.frbcat.org)$.}. The DM distribution of the observed FRBs is sufficient to constrain BBH merger models. For example, only $\sim 100$ GW-less FRBs with $\overline{DM}<500 {\rm pc~cm^{-3}}$ in the aLIGO era can reach the constraint that the fraction of FRBs from BBH mergers is less than $10\%$. Since the aLIGO horizon for BNS merger is small, it would take a long time to reach the desired sample to constrain the BNS-origin FRB models. This process will speed up in the LIGO A+ and ET era.

We also proposed a method to constrain the charge of BHs in BBH merger systems. With the fraction of no-BBH-merger FRBs constrained to below $\overline{f}_{\rm BBH}<0.93^{+1.03}_{-0.50}\%$ for relavant FRBs whose DM values fall into the BBH merger horizon, one can start to place a limit on the BH charge for the first time, as shown in Eq.(\ref{eq:q-constraints}) and (\ref{eq:F}).

Different BNS FRB models \citep{Totani2013,Zhang2014,Wang2016} predict FRBs to occur in different merging phases, so that one should search BNS-FRB associations with different time offsets. These different models also predict different degrees of beaming angles (e.g. for FRBs produced during and after the merger, only a small fraction of the solid angle is transparent for radio waves). Our constraints on the validity of these models should properly consider the beaming correction of the observed event rate of FRBs.

\acknowledgments
This work is supported by the National Natural Science Foundation of China under Grant No. 11722324, 11690024,  11603003, 11633001, the Strategic Priority Research Program of the Chinese Academy of Sciences, Grant No. XDB23040100 and the Fundamental Research Funds for the Central Universities.

\end{document}